\documentclass[journal=jacsat,manuscript=article]{achemso}
\usepackage[version=3]{mhchem} 
\usepackage[T1]{fontenc}       

\usepackage{mathrsfs}
\usepackage{graphicx}
\usepackage{bm}        
\usepackage{amssymb}   
\usepackage{amsmath}
\usepackage{color}
\usepackage{subfigure}
\usepackage{url}

\usepackage{siunitx}

\usepackage{color}
\definecolor{mygreen}{rgb}{0, 0.6, 0}

\def\bea{\begin{eqnarray}}
\def\eea{\end{eqnarray}}
\def\be{\begin{equation}}
\def\ee{\end{equation}}

\usepackage[colorlinks=true,linkcolor=blue,citecolor=blue,urlcolor=blue]{hyperref}

\author{Amir Khosravanizadeh}
\email{khosravani@iasbs.ac.ir}
\affiliation{Department of Physics, Institute for Advanced Studies in
Basic Sciences (IASBS), Zanjan 45137-66731, Iran}

\author{Pierre Sens}
\email{pierre.sens@curie.fr}
\affiliation{Institut Curie, PSL Research University, CNRS, UMR 168, 26 rue d'Ulm, F-75005 Paris, France}

\author{Farshid Mohammad-Rafiee}
\affiliation{Department of Physics, Institute for Advanced Studies in
Basic Sciences (IASBS), Zanjan 45137-66731, Iran}
\altaffiliation{Research Center for Basic Sciences \& Modern Technologies $(\text{RBST})$, Institute for Advanced Studies in Basic Sciences (IASBS), Zanjan 45137-66731, Iran}

\email{farshid@iasbs.ac.ir}

\date{\today}

\title[An \textsf{achemso} demo]
  {Receptor-Mediated Endocytosis of a Cylindrical Nanoparticle in the Presence of Cytoskeleton Substrate}

\abbreviations{IR,NMR,UV}
\keywords{Receptor-Mediated Endocytosis, Cylindrical Nanoparticle, Molecular Dynamics Simulations, Cytoskeleton, Asymmetric Wrapping}

\begin{document}





\begin{abstract}
Internalization of particles by cells plays a crucial role for adsorbing nutrients and fighting infection. Endocytosis is one of the most important mechanisms of the particles uptake which encompass multiple pathways. Although endocytosis is a complex mechanism involving biochemical signaling and active force generation, the energetic cost associated to the large deformations of the cell membrane wrapping around the foreign particle is an important factor controlling this process, which can be studied using quantitative physical models. Of particular interest is the competition between membrane - cytoskeleton and membrane - target adhesion. Here, we explore the wrapping of a lipid membrane around a long cylindrical object in the presence of a substrate mimicking the cytoskeleton. Using discretization of the Helfrich elastic energy that accounts for the membrane bending rigidity and surface tension, we obtain a wrapping phase diagram as a function of the membrane-cytoskeleton and the membrane-target adhesion energy that includes unwrapped, partially wrapped and fully wrapped states. We provide an analytical expression for the boundary between the different regimes. While the transition to partial wrapping is independent of membrane tension, the transition to full wrapping is very much influenced by membrane tension. We also show that target wrapping may proceed in an asymmetric fashion in the full wrapping regime.
\end{abstract}


\section{Introduction}
Lipid bilayer membranes are one of the most important components of living cells that separate individual organelles of cells from each other or cells from their surrounding environments. 
Regulation of exchange across the cell membrane is fundamental for the intake of nutrients and the interaction of a cell with its environment. The translocation of particles across  the cell membrane depends on the size of particles (ranging from sub-nanometer to a few microns) \cite{lodish}. While sub-nanometer particles like ions can translocate through the bilayer by direct diffusion or via membrane channels \cite{orsi2010permeability}, larger objects enter the cell by endocytosis or phagocytosis, during which the cell membrane experiences large conformational deformations in order to engulf the target \cite{swanson2008shaping}. Here we focus on the internalization of nanoparticles (NPs), which play an important role in biomedical fields such as chemotherapy, bioimaging, biosensing, and drug and gene delivery \cite{xia2008nanomaterials,weissleder2006molecular,nel2009understanding,allen2004drug,whitehead2009knocking,peer2007nanocarriers}, but can also create
cytotoxicity that may cause damage to the cells \cite{murphy2008gold}. In this regard, the mechanisms of NPs uptake have been extensively studied during the last decade \cite{conner2003regulated,chen2013internalization,
yang2010computer}.
 
Experimental studies indicate that direct penetration through transient membrane pores and receptor-mediated endocytosis are
two main pathways for NPs internalization \cite{verma2008surface,wang2012cellular,ruan2007imaging,chithrani2006determining,
shi2011cell}. 
Endocytosis of NPs encompass multiple pathways, from clathrin-mediated and caveolae-mediated endocytosis, which mostly involves the formation of membrane protein coats, to phagocytosis and  macropinocytosis, for which the cytoskeleton plays an important part and could be accompanied by a symmetry breaking in the membrane protrusion \cite{conner2003regulated}.

Beside the experimental studies, various theoretical models and computer simulations have been performed for exploring details of NPs internalization mechanisms \cite{gao2005mechanics,
vacha2011receptor,li2012surface,li2008computational,yi2011cellular,yue2013molecular,ding2012role,yuan2010variable,smith2007designing,deserno2004elastic}. Membranes can be modeled with a broad range of approaches at different details and time scales. Atomistic simulations give a molecular description of the process, but are limited to short length and time scales
\cite{tieleman1997computer,zubrzycki2000molecular}. Analytical theories and triangulated membrane simulations can yield the shape and phase diagrams of membranes at large length scales, but do not explain what happens at the molecular level \cite{seifert1997configurations,kumar2001budding}. In  middle of this range, coarse-grained models have been developed to reduce length and time scales limitations, while preserve main properties of membranes \cite{goetz1998computer,noguchi2001self,farago2003water,
cooke2005solvent,yuan2010one}.

For direct penetration, studies show size,\cite{roiter2008interaction,ding2012designing} shape\cite{ding2012designing,yang2010computer} and surface properties\cite{verma2008surface,li2012surface,li2008computational} of NPs play crucial roles in internalization of particles. In the case of receptor-mediated endocytosis, theories, simulations and experiments show that engulfment of NPs is determined by elastic properties of membrane such as its tension and bending rigidity,\cite{yi2011cellular,lipowsky1995structure,hashemi2014regulation,deserno2004elastic,
bahrami2014wrapping,seifert1997configurations} the size\cite{chithrani2006determining,zhao2011interaction,hashemi2014regulation,
deserno2004elastic,chaudhuri2011effect,bahrami2014wrapping,zhang2009size} and shape\cite{chithrani2006determining,shi2011cell,gao2005mechanics,
vacha2011receptor,chen2016shape,bahrami2014wrapping,bahrami2013orientational,dasgupta2014membrane} of the NP target, its surface properties,\cite{zhao2011interaction,yi2011cellular,yue2013molecular,ding2012role} and binding strength of ligands-receptors between NPs and biomembrane\cite{yi2011cellular,hashemi2014regulation,yuan2010variable,bahrami2014wrapping,yue2012cooperative}

 An important aspect of cell membrane mechanics that has been overlooked in existing models of NPs internalization is the presence of an actin-based network underneath the plasma membrane, called cytoskeleton cortex \cite{bray}, that strongly influences the membrane's ability to deform \cite{hashemi2014regulation}.
The aim of the present study is to assess the impact of membrane-cortex attachment on the internalization of a single NP. We concentrate on simulating uptake of an infinite long cylindrical NP by introducing a 2D molecular dynamics model which is constructed by discretization
of the Helfritch energy. In the Results section, we present the simulation results in a phase diagram including three different regimes (unwrapped, partially, and fully wrapped) and two phase transitions. The result section continues with the analytical derivation of the different transitions, and a discussion of the symmetry breaking process accompanying the full wrapping transition, which bears strong relevance to the macropinocytosis process.

\section{RESULTS AND DISCUSSION}
\begin{figure}
\centering
\includegraphics[width=\columnwidth ]{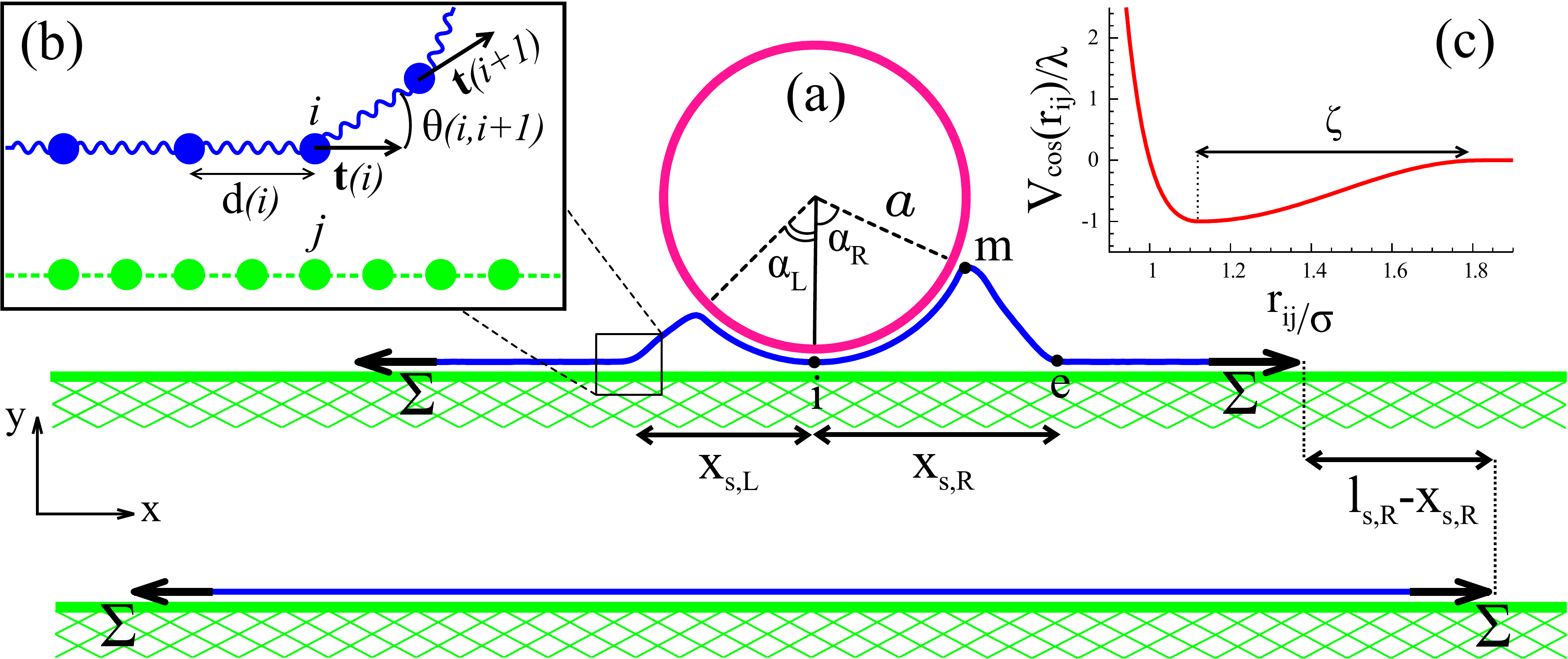}
\caption{(a) The schematic picture of an infinite long cylindrical NP (pink) engulfed by a flat membrane (blue) attached to an non-deformable cytoskeleton substrate (green). The surface tension of the membrane is denoted by $\Sigma$. (b) a closer view at the bead-spring membrane model; (c) The membrane adhesion potential $V_{cos}(r_{ij})$, see Eq. (\ref{eq:dt}), for modeling ligand-receptor interactions. }
\label{fig1}
\end{figure}

The endocytosis of an NP can be significantly affected by different factors such as the surface tension and the bending rigidity of the membrane, the NP size, and the membrane-cytoskeleton and the membrane-NP adhesion energy. The simulations presented in this work are based on a coarse-grained description of the membrane, shown in  Fig. \ref{fig1}. The cell membrane (in blue), adhered to a flat cytoskeleton cortex sheet (in green) is deformed owing to its interaction (adhesion) with an external particle (pink), a cylinder of radius $a$. Membrane wrapping around the target results from a competition between membrane-target and membrane-cortex adhesion, and is modulated by the membrane elasticity. It is characterized by the wrapping angles $\alpha_R$ and $\alpha_L$ at the right and left side of the particle, respectively.

The adhesion energy difference with respect to the reference state where the membrane is flat and fully adhered to the cytoskeleton is,
per unit length of the cylinder:
\begin{equation}
\Delta E_{ad}=-\omega a \alpha +\omega_s (l_{s,L} + l_{s,R}),
\label{e_ad}
\end{equation}
where $\omega$ and $\omega_s$ indicate the membrane-target and membrane-cytoskeleton adhesion energies per unit area, respectively. The total wrapping angle is $\alpha = \alpha_L + \alpha_R$ and  $l_{s,L}$ and $l_{s,R}$ denote the left and right membrane contour length detached from the cysotkeleton (the contour length between the points $i$ and $e$ in Fig. \ref{fig1} (a)). In principle the engulfment can happen in an asymmetric way and therefore $l_{s,L}$ (respectively $\alpha_L$) can be different from $l_{s,R}$ (respectively $\alpha_R$). 

The elastic energy of the deformed membrane can be calculated using the Helfrich energy \cite{helfrich1973elastic} as
\begin{equation}
E_H=\frac{1}{2}\kappa\int_a \left( C_1+C_2 \right)^{2}da+\Sigma\int_a da,
\end{equation}
where $\kappa$ and $\Sigma$ denote the bending rigidity and the surface tension of the membrane, respectively, and $C_1$ and $C_2$ are the local principle membrane curvatures. Thanks to the translational invariance along the cylinder axis, a single tangent vector $\hat{t}(s)$ defined at each point $s$ of the membrane suffices to fully characterize its shape. The total deformation energy per unit length of the cylinder is written as
\begin{equation}
E_H=\frac{1}{2}\kappa\int_s \left[ \partial_s \hat{t}(s) \right]^{2} ds+\Sigma\int_s ds,
\label{one_dimension}
\end{equation}
where $\partial_s \hat{t}(s)$ denotes differentiation with respect to $s$.
For the computer simulations, the membrane is discretised into a chain of beads and springs, and the cylinder and the cytoskeleton are the collections of beads, as well. The membrane-cytoskeleton and membrane-NP ligand-receptor interactions are compensated by an attraction potential between the beads (see Fig. \ref{fig1} (c) and Method section) and adding a lateral force at the edge of the membrane reproduces the effect of the membrane surface tension $\Sigma$. We set the values of $\kappa=20 \, \varepsilon$, $a=50 \, \sigma$, where $\epsilon$ and $\sigma$ are the typical energy and length scales of the simulation (see the Method section), and investigate the wrapping process as a function of the cortex adhesion energy $\omega_s$ and the target adhesion energy $\omega$ for two values of the membrane tension $\Sigma=1.5 \, \varepsilon/\sigma^{2}$ and $\Sigma=\omega_s$.

\begin{figure}
\centering
\includegraphics[width=1.0\columnwidth]{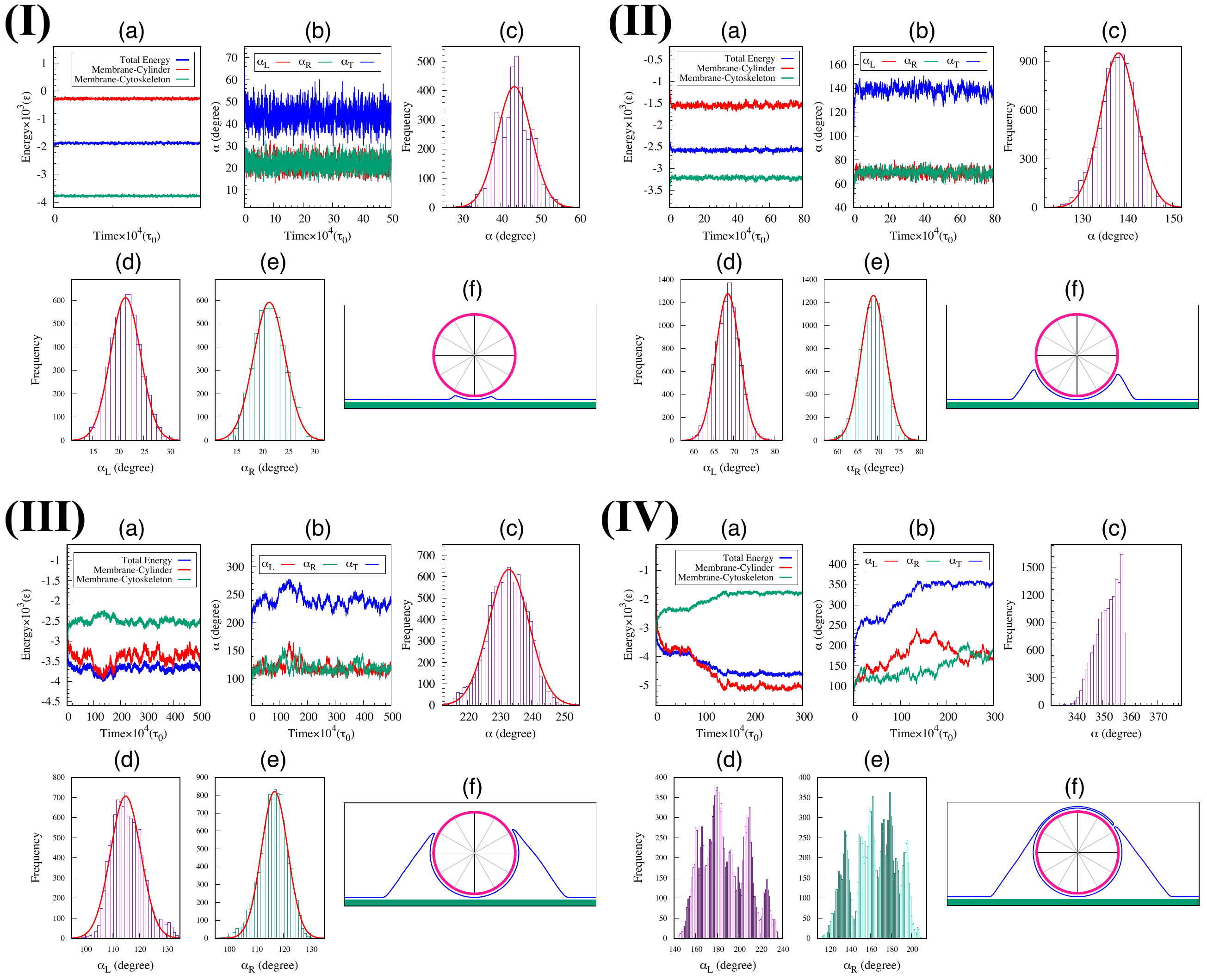}
\caption{Simulation results for different values of the membrane-target adhesion energy $\omega=7.31, ~ 12.73, ~ 16.08, ~ 16.19 \, (\varepsilon/\sigma^2)$ from plots (I) to (IV), with $\omega_s = \Sigma= 3.95 \, (\varepsilon/\sigma^2)$. Time evolution of (a) the total energy (blue), membrane-cytoskeleton adhesion energy (green) and membrane-cylinder adhesion energy (red); 
(b) the total wrapping angle (blue), left side angle (red) and right side angle (green). Panels (c), (d), and (e) represent the steady-state distribution of the total wrapping angle, the left side angle and the right side angle, respectively. In (f) typical snapshots of the membrane conformation in the equilibrium state are shown. While in partial wrapped regimes (I to III), the engulfment of the particle is symmetric, whereas in sufficient large $\omega$ (panel IV), the engulfment can be asymmetric. (Movie. 1 in the Supporting Information)}
\label{fig2}
\end{figure}

In our simulations, the system reaches equilibrium after a sufficient number of MD steps. The results of simulations are shown in Fig. \ref{fig2}. Panels (a) show the total energy of the system, the membrane-cylinder adhesion energy and the membrane-cytoskeleton adhesion energy as a function of time. Panels (b) show the total engulfment angle, $\alpha$ and the left and right angles $\alpha_L$, and $\alpha_R$ as a function of time. In panels (c)--(e), the distribution of the total engulfment angle, $\alpha$, the wrapping angle from the left side of the object, $\alpha_L$, and right side, $\alpha_R$ are shown, when the system becomes equilibrated. The plots (I) to (IV) correspond to $\omega = 7.31, ~ 12.73, ~ 16.08, ~ 16.19 \, (\varepsilon/\sigma^2)$, respectively, with $\omega_s = 3.95 \, (\varepsilon/\sigma^2) $ and $\Sigma = \omega_s $. Considering $\sigma=3 \, nm$ as the thickness of a typical lipid bilayer, and $\varepsilon=1 \,k_BT$, the mentioned values are corresponding to $\kappa=60 \, k_BT$, $\omega_s = \Sigma= 0.44 \, (k_BT/nm^2)$, and $\omega = 0.81, ~ 1.41, ~ 1.79, ~ 1.79 \, (k_BT/nm^2)$.

\begin{figure}
\centering
\begin{tabular}{cc}

\includegraphics[width=0.5\columnwidth]{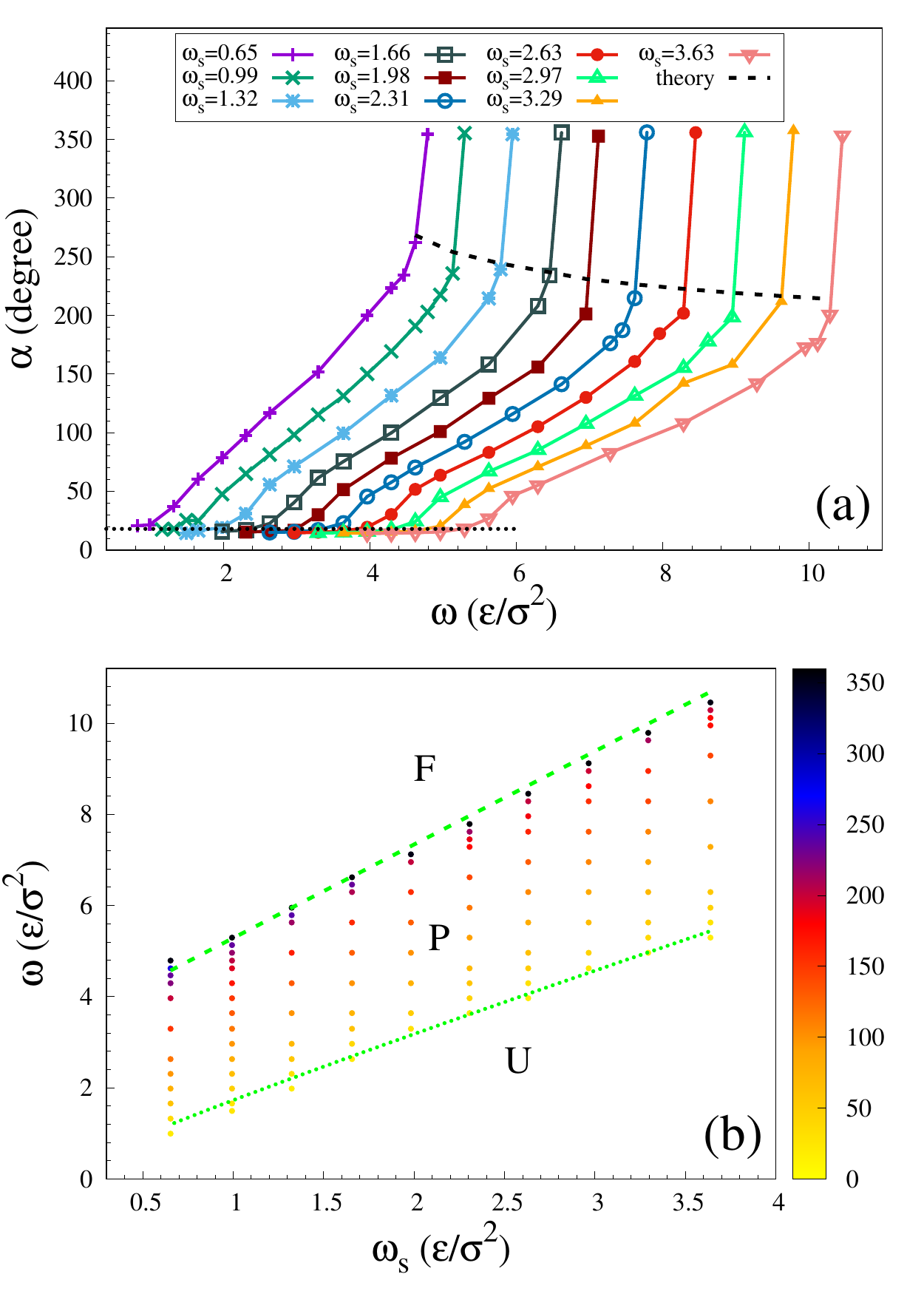}&
\includegraphics[width=0.5\columnwidth]{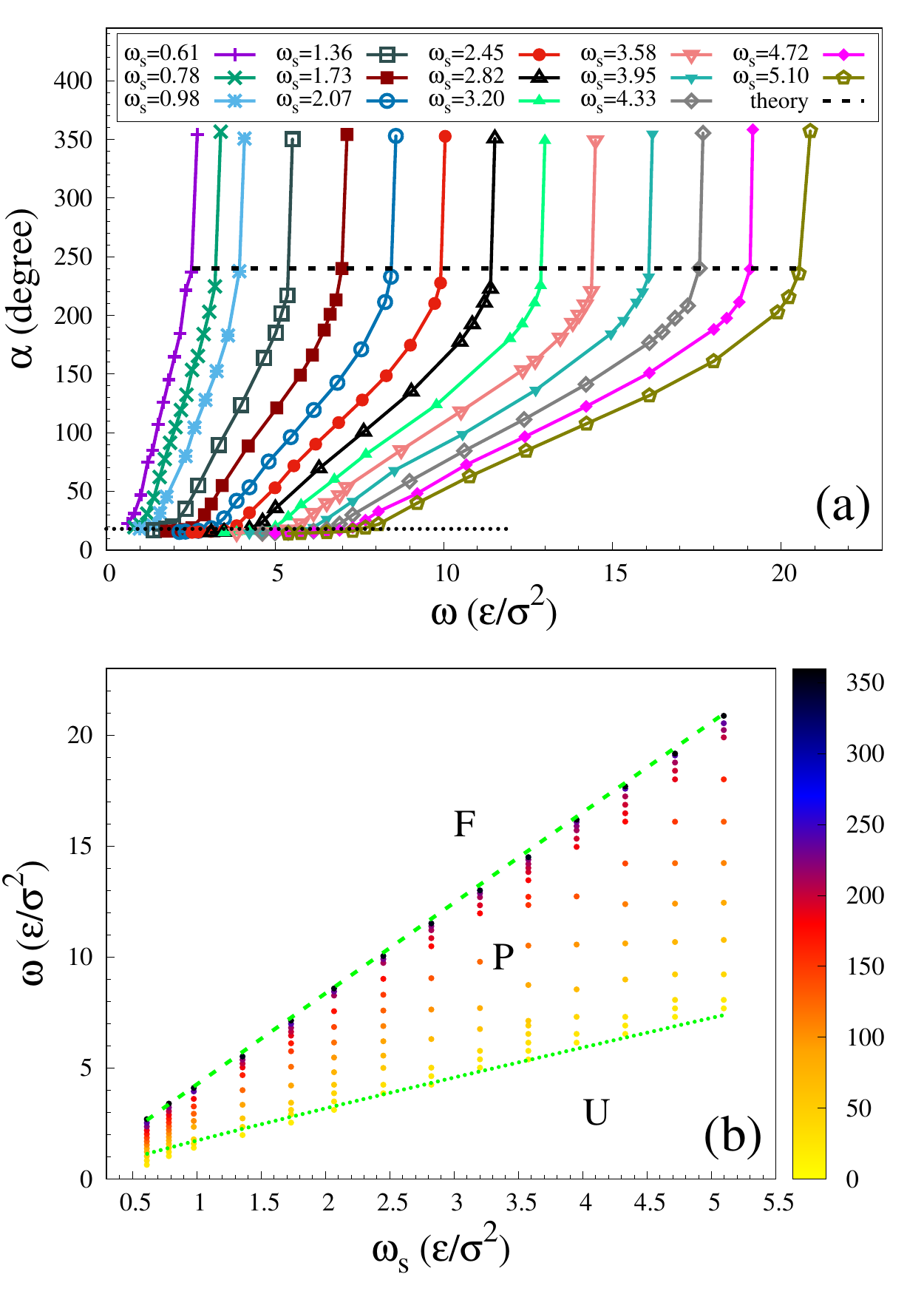}
\end{tabular}
\caption{(a) Variation of the wrapping angle with $\omega$ for different values of $\omega_s$ in the case of $\Sigma=1.5 \, \varepsilon/\sigma^{2}$ (left) and $\Sigma=\omega_s$ (right). The dotted line corresponds to the transition to a partially wrapped state while the dashed line shows the theoretical value of the wrapping angle ($\theta_f$) to a fully wrapped state, see Eq. (\ref{teta_f1}). (b) Wrapping phase diagram of the cylindrical NP in the phase space of $\omega_{s}$ and $\omega$ for the same values of membrane tension $\Sigma$. The color bar indicates the extent of engulfment angle (in degrees).
There are three different regimes: Fully wrapped (F), partially wrapped (P), and unwrapped (U). The boundaries between the different regimes given by Eq. (\ref{omega-u1}) (dotted line) and Eq. (\ref{omega-f1}) (dashed line). }
\label{fig3}
\end{figure}

The equilibrium wrapping angle, $\alpha$ is shown in Fig. \ref{fig3} (a) as a function of the target adhesion energy $\omega$ and the cytoskeleton adhesion energy $\omega_s$, for two values of membrane tension: $\Sigma=1.5 \, \varepsilon/\sigma^{2}$, and $\Sigma=\omega_s$. The phase diagram of the system as a function of $\omega$ and $\omega_s$ is shown for the two values of $\Sigma$ in Fig. \ref{fig3} (b). For small values of the target adhesion energy $\omega$, the wrapping angle is very small ($\lesssim \ang{18}$). This is the unwrapped regime ($U$). The non-zero value of the wrapping angle in this regime is due to the finite range of the adhesion interaction in Eq. (\ref{eq:dt}) (see Fig. \ref{fig1} (c)). For values of $\omega$ exceeding a threshold $\omega_u$, that increases with $\omega_s$, the wrapping angle continuously increases with $\omega$ - see Fig. \ref{fig2} (I) to (III). This is the partially wrapped regime (P) and it is nearly symmetric in the left and right. The transition from the unwrapped to the partial wrapped state ($U-P$ transition) can be understood analytically by performing an expansion of the membrane deformation energy for small wrapping angle $\alpha\ll 1$. As it explained in the Supporting Information section (a), by expanding the energy of the system to third order of wrapping angle and an energy minimization, $\bar{\omega}_u$ can be calculated as,
\begin{equation}
\bar{\omega}_u=\left( 1+\sqrt{\bar{\omega}_s} \right)^2+\frac{4}{9}\left[ \left( 1+3\sqrt{\bar{\omega}_s} \right)^{3/2}-1  \right],
\label{omega-u1} 
\end{equation}
where the dimensionless quantities are defined as
\begin{equation}
\bar{\omega}\equiv\frac{2\omega a^2}{\kappa}; \hspace{5mm} \bar{\omega}_s\equiv\frac{2\omega_s a^2}{\kappa}; \hspace{5mm} \bar{\Sigma}\equiv\frac{2\Sigma a^2}{\kappa}.
\label{dimensionless1}
\end{equation}
This transition is indicated by a dotted line that separates the unwrapping region from the partial wrapping region in Fig. \ref{fig3} (b).

Beyond yet another critical value $\omega_f$, which also increases with $\omega_s$, an abrupt transition occurs from the partially wrapped to a fully wrapped state ($F$), where the cylinder is totally engulfed by the membrane. In this regime, the left and right angles fluctuate over time, always adding to a total wrapping angle $\alpha\simeq\ang{360}$ - see Fig. \ref{fig2} (IV) and Movie. 1 in the Supporting Information. The transition to fully wrapped state can be understood by exploring the behavior of a generalized force acting on the membrane which can be extracted from the energy changes associated to infinitesimal membrane displacements. As is shown in Supporting Information (section (b)), there exists a
transition angle $\theta_f$, given by
\begin{equation}
\tan \theta_f = - \frac{\sqrt{{\bar{\omega}_s}^2+2\bar{\Sigma}\bar{\omega}_s}}{\bar{\Sigma}},
\label{teta_f1}
\end{equation}
beyond which an abrupt transition to full wrapping occurs. It should be noted that $\theta$ has been replaced instead of $\alpha_L$ or $\alpha_R$.
In the case of $\Sigma=\omega_s$,
$\tan \theta_f = - \sqrt{3}$ and $\theta_f=\frac{2 \pi}{3}$. The dashed line in Fig. \ref{fig3} (a) represents the analytical values of the wrapping angle ($2\theta_f$) at the transition point. The transition to full wrapping occurs when the membrane-cylinder binding energy reaches the value:
\begin{equation}
\bar{\omega}_f=\left( 1+\sqrt{2\left( \bar{\Sigma}+\bar{\omega}_s \right) }\right)^2
\label{omega-f1}
\end{equation}
This transition is shown as a dashed line in Fig. \ref{fig3} (b).

\subsection{ Energy of the fully wrapped state} 
\begin{figure}
\centering
\includegraphics[width=1.0\columnwidth]{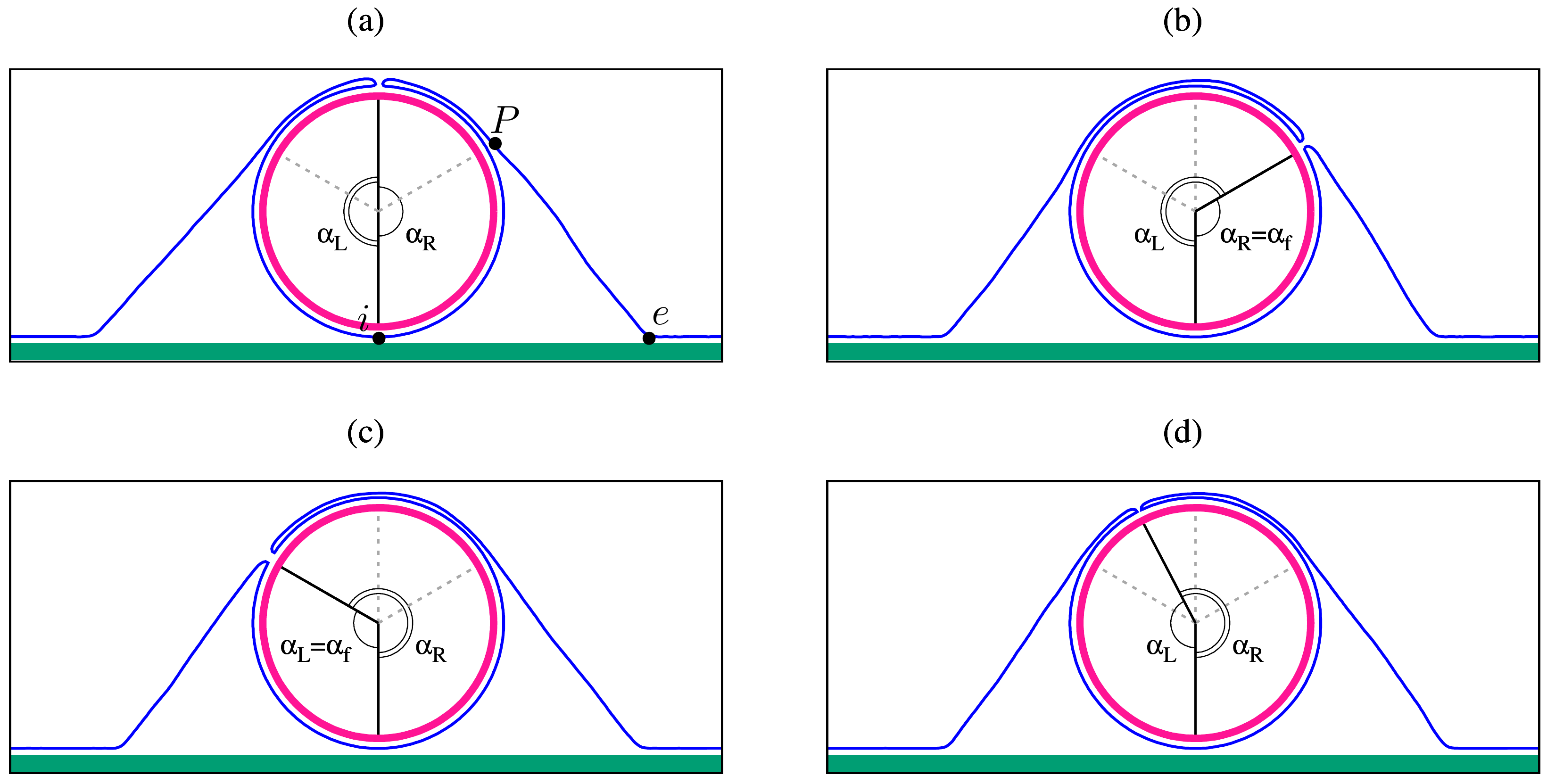}
\caption{Typical snapshots of different wrapping states for $\Sigma=\omega_s$. The parameters of the plots are: (a) $\omega_s=2.45$ and $\omega=10.04$, (b) $\omega_s=2.82$ and $\omega=11.52$, (c) $\omega_s=3.58$ and $\omega=14.50$, and (d) $\omega_s=3.95$ and $\omega=16.19$.}
\label{asymmetric}
\end{figure}

Our simulation results show that, while the partially wrapped states show left-right symmetry, the fully wrapped state displays strong fluctuations with respect to the left and right wrapping angles (Fig. \ref{fig2} (IV) and Movie. 1 in the Supporting Information). This is further illustrated in Fig. \ref{asymmetric}, where typical snapshots of fully wrapped states are shown for different sets of parameters. As we now show, this is a consequence of the fact that within some bounds, the energy of the fully wrapped configuration only depends on the total wrapping angle.

Inspection of the membrane shape in the full wrapped regime (see Fig. \ref{asymmetric}) shows that the left and right membrane halves can each be decomposed into four distinct membrane segments: {\em (i)}: the inner cap, extending from $\alpha=0$ to an angle $\alpha_i$ ($i=L,\ R$ for the left and right halves),  which is in direct contact with the target; {\em (ii)}: the outer cap, which is not touching the target, but whose shape is fixed by the target shape through steric repulsion, that extend between $\alpha_i$ and a particular angle $\alpha_{\rm f}$
(the angle of the point $P$, where the membrane leaves the target, in Fig. \ref{asymmetric}); {\em (iii)}: a free membrane tail, not adhered to any substrate (between points $P$ and $e$ in Fig. \ref{asymmetric}); and {\em (iv)}: the flat part attached to the cytoskeleton.  Considering that the free tail is to a very good approximation straight (see Fig. \ref{asymmetric}), and neglecting the contribution of the bending energy at points $P$ and $e$, the energy difference (with respect to the reference state) of the left and right parts of the membrane can be written as ($i=L,\ R$ for the left and right halves):
 \begin{eqnarray}
\Delta E_i=\frac{\kappa}{2a}(2\alpha_i-\alpha_{\rm f})-\omega a\alpha_i+\omega_s l_{s,i}+\Sigma(l_{s,i}-X_{s})
\label{E_asymmetric}
\end{eqnarray}
where $l_{s,i}$ is the contour length of the deformed membrane (segments {\em (i)} to {\em (iii)}) and $X_{s}$ is the location of point $e$), see Fig. \ref{fig1}. Geometric consideration yield $l_{s,i}=(2\alpha_i-\alpha_{\rm f})a+D$ and $X_s \simeq D$, where the length of the free tail (segment {\em (iii)}) is $D=a\tan{\frac{\alpha_{\rm f}}{2}}$.  

The total energy $E=E_R+E_L$ can be written:
\begin{eqnarray}
\Delta E=\frac{\kappa}{2a}\left[2(\alpha_R+\alpha_L)\left((1-\frac{\bar\omega}{2}+\omega_s+\bar\Sigma\right)\right.\cr
\left.-2\alpha_{\rm f}(1+\bar\omega_s+\bar\Sigma)+2\bar\omega_s\tan{\frac{\alpha_{\rm f}}{2}}\right]
\end{eqnarray}
This energy does not individually depend on $\alpha_R$ and $\alpha_L$, but only on the total wrapping angle $\alpha=\alpha_R+\alpha_L$, which is fixed to $\alpha=2\pi$ in the fully wrapped regime. This explains the degeneracy of equilibrium shapes illustrated by Fig. \ref{asymmetric}. Minimisation of this energy with respect to the ``tail angle'' $\alpha_{\rm f}$ yields:
\begin{equation}
\cos{\frac{\alpha_{\rm f}}{2}}=\sqrt{\frac{\bar\omega_s}{ 2\left(1+\bar\omega_s+\bar\Sigma \right)} }
\label{theta_t}
\end{equation}
Interestingly, this angle is equal to the critical wrapping angle at the full wrapping transition (Eq. (\ref{teta_f1})) if $\bar\omega_s\gg1$ and $\bar\Sigma\gg 1$, and is equal to $\pi/2$ when $\bar\omega_s\gg\bar\Sigma,1$.

\section{Conclusion}

In this paper, we introduced a new coarse-grained model for simulating lipid membranes in two dimension which is constructed, by discretization of the Helfritch energy. Using this model we study the engulfment of a cylindrical nanoparticle by a lipid membrane adhered to a planar substrate mimicking the cortical cytoskeleton. The competition between the membrane adhesion energy with the target (characterized by the parameter $\omega$) and with the cytoskeleton (parameter $\omega_s$) defines three distinct regimes of engulfment, separated by two phase transitions (see Fig. \ref{fig3}). The target remains unwrapped (U) by the membrane if $\omega<\omega_u$, given by Eq. (\ref{omega-u1}). The target is fully engulfed (F) by the membrane if $\omega>\omega_f$ given by Eq. (\ref{omega-f1}). In between: $\omega_u<\omega<\omega_s$, the target is partially engulfed (P). Both critical target adhesion energies $\omega_u$ and $\omega_f$ are increasing functions of the cytoskeleton adhesion energy $\omega_s$ and of the membrane bending rigidity $\kappa$. The full wrapping threshold $\omega_f$ also increases with the membrane tension, while the partially wrapping threshold is insensitive to membrane tension. The wrapping angle $\alpha$ smoothly increases with $\omega$ in the partially wrapped region, to reach a critical value $\alpha=2\theta_f$, given by Eq. (\ref{teta_f1}), when $\omega=\omega_f$. The wrapping angle abruptly jump to $\alpha=2\pi$ at the full wrapping transition. 
The critical angle is $\theta_f=\pi/2$ for a membrane with vanishing surface tension ($\Sigma=0$), and is equal to $\theta_f=2\pi/3$ if the surface tension matches the cortex binding energy ($\Sigma=\omega_s$).  

The membrane bending rigidity is of order $\kappa\simeq 20 \, k_BT$ \cite{lipowsky1995structure}, while the surface tension of the cell membrane can vary several orders of magnitude: $\Sigma \simeq 10^{-6}-10^{-3} \, N/m$\cite{morris2001cell}. For a cylindrical nanoparticle of radius $a=100$nm, the dimensionless tension (Eq. (\ref{dimensionless1})) is within the range $\bar\Sigma=0.25-250$. The cytoskeleton adhesion energy can be estimated to be of order $k_BT$ per membrane/cortex binding site giving $\omega_s\simeq 10^{-4}N/m$ and $\bar\omega_s=25$. With these values, the transition to partial wrapping occurs for $\omega_u\simeq 2.56 \, \omega_s$. We find an almost direct transition to a full wrapping situation ($\omega_f\simeq 2.6\omega_s\gtrsim\omega_u$) for vanishing membrane tension ($\Sigma=0$), and a fairly extended partially wrapped regime ($\omega_f\simeq 11.3 \, \omega_s \gg\omega_u$) for large membrane tension ($\bar\Sigma=100$).

In the full wrapping regime, the system is degenerate, which means that the left and right wrapping angle may fluctuate between $\alpha_{\rm f}$ and $2\pi-\alpha_{\rm f}$, with $\alpha_{\rm f}$ given by Eq. (\ref{theta_t}). We thus predict that asymmetric membrane protrusions should be commonly observed in this regime. Interestingly, macropinocytosis, an important pathway for particle internalization in cells, is characterized by asymmetric membrane protrusions \cite{conner2003regulated}. While this process clearly involves multiple complex processes related to signaling and actin polymerization, our analysis provides a physical explanation for this asymmetry. 

The present model is relatively simple and tunable, and can be extended to include additional mechanical effects relevant to cellular membranes, such as actin polymerisation underneath the membrane, or the formation of protein coat. Such model can easily be used to study other cellular phenomena such as clathrin-mediated endocytosis or cell crawling.

\section{MODEL AND SIMULATION METHOD}
For the computer simulations, the membrane is discretised into a chain of beads and springs with $\bm{t}\left(i\right)$ the direction of $i$th spring (with $\left| \bm{t}\left(i\right)\right|=1$) and $\theta(i,i+1)$ the angle between neighbouring springs (see Fig. \ref{fig1} (b)). The discretized bending energy reads \cite{kierfeld2004stretching}:
\bea
E_B &=& \frac{1}{2}\kappa \sum_{i=1}^{N-2} \left[ \frac{\bm{t}\left(i+1\right)-\bm{t}\left(i\right)}{d_0}\right]^{2}d_0 \nonumber \\
&=& \frac{\kappa}{d_0} \sum_{i=1}^{N-2} \left[1-\cos\theta \left(i,i+1\right)  \right],
\label{chains}
\eea
which defines the effective bending stiffness $\kappa_e=\frac{\kappa}{d_0}$. The harmonic spring potential energy is:
\begin{equation}
E_{spring}=\frac{1}{2} \Lambda \sum_{i=1}^{N-1} \left[ {d\left(i\right)-d_0}\right]^{2},\label{spring_potential}
\end{equation}
where $\Lambda$ is the stiffness of the springs, $d\left(i\right)$ is the bond length, and $d_0$ is the equilibrium bond length.
Adding a lateral force at the edge of the membrane reproduces the effect of the membrane tension $\Sigma$. 

Excluded volume interactions between the membrane beads are implemented using the Weeks-Chandler-Andersen potential
\begin{align}
V_{LJ}\left(r_{ij}\right)=\left\lbrace 
\begin{array}{cc}
4\varepsilon \left[ \left( \frac{\sigma}{r_{ij}}\right) ^{12}-\left( \frac{\sigma}{r_{ij}}\right) ^{6}+\frac{1}{4} \right],  & r_{ij}\leq 2^{1/6}\sigma\\
0, & r_{ij}> 2^{1/6}\sigma,
\end{array} \right.
\end{align}
where $\varepsilon$ and $\sigma$ are our unit energy and length scale, respectively, and $r_{ij}$ is the distance between the $i$th and $j$th beads. The diameter of the membrane monomers is $R=d_0 \doteq 1\sigma$. In our simulations, the membrane is constructed by $1000$ monomers, and is allowed to move only in $x-y$ plane, while the cytoskeleton substrate and the target consist of beads fixed in space. The cytoskeleton is made of
$2600$ monomers laid underneath the membrane, and the cylindrical NP is assembled from $628$ monomers positioned on top of the membrane. The cylinder is placed in the center of the membrane and its radius is $a= 50 \, \sigma$. The distance between monomers in the cytoskeleton and the NP is chosen to be $0.5 \, \sigma$ in order to simplify the reptation of the membrane on in these structures.

Both the membrane-cytoskeleton and the membrane-NP interactions are modeled with the following potential,
\begin{eqnarray}
V_{cos}\left(r_{ij}\right) = 
  \begin{cases}
 4\lambda_{k} \left[ \left( \frac{\sigma}{r_{ij}}\right) ^{12}-\left( \frac{\sigma}{r_{ij}}\right) ^{6} \right],  &\quad  r_{ij} < 2^{1/6}\sigma \\
-\lambda_{k} \cos^{2}\left[\frac{\pi}{2\zeta}\left(r_{ij}-2^{1/6}\sigma\right)\right] , &\quad 2^{1/6}\sigma \leq r_{ij} \leq 2^{1/6}\sigma+\zeta \\
0,  &\quad  r_{ij} > 2^{1/6}\sigma+\zeta,
  \end{cases}
  \label{eq:dt}
\end{eqnarray}
where $\lambda_k$ denotes the strength of the ligand-receptor interactions ($k=1$ corresponds to the membrane-NP, and $k=2$ corresponds to the membrane-cytoskeleton ligand-receptor interactions) with dimension of energy. The interaction potential (see Figure \ref{fig1} (c)) smoothly decays to zero for $r>2^{1/6} \, \sigma$ and the attraction tail depends on the value of $\zeta$. The values of the average adhesion energy per unit length $\sigma$ between the membrane and the cytoskeleton ($\omega_s$) and between the membrane and the cylinder ($\omega$) can be tuned by varying $\lambda_1$ and $\lambda_2$.

Our Molecular Dynamics (MD) simulations were performed at the constant temperature $T=1.0 \, \varepsilon/k_B$, with the Langevin thermostat, and using ESPResSo \cite{espresso}. The time step in the Verlet algorithm and the damping constant in the Langevin thermostat were set $\delta t=0.01 \,  \tau_0$ and $\Gamma=\tau^{-1}_0$, respectively, which $\tau_0=\sqrt{\frac{m\sigma^{2}}{\varepsilon}}$ is the MD time scale and $m$ is the monomer mass. Also, it is worth to mention that the values of the spring stiffness ($\Lambda$) and of the (short) range of the interaction potentials ($\zeta$) do not affect our results. We fix $\Lambda=5000 \, \varepsilon/\sigma^{2}$ and $\zeta=0.5 \, \sigma$ (for $\Sigma=1.5 \, \varepsilon/\sigma^2$) or $\zeta=0.7 \, \sigma$ (for $\Sigma=\omega_s$) to optimize the convergence of the simulation.




\section{Supporting Information}

\subsection{(a) Unwrapped-partial wrapped transition}
The transition from the unwrapped to the partial wrapped state ($U-P$ transition) can be understood analytically by performing an expansion of the membrane deformation energy for small wrapping angle $\alpha\ll 1$. In small angles, the wrapping angle is nearly symmetric on the left and right, so we can write $\alpha_R=\alpha_L=\theta$ and $l_{s,R}=l_{s,L}=l_s$.  
In general, the total energy of the system, which includes contribution from the bending energy, the surface tension, and the gain of membrane-target adhesion and the loss of membrane-cytoskeleton adhesion, can be divided in two main parts: the cap (the part of the membrane wrapped onto the target) and the tail (the free part of the membrane, in contact with neither the target nor the cytoskeleton). Their contributions to the energy difference (per unit cylinder length) with respect to the reference state fully adhered to the cytoskeleton are:
\begin{equation}
\Delta E_{cap}=\frac{\kappa \theta}{a}+2\Sigma \theta a \left( 1- \frac{\sin \theta}{\theta} \right)+2(\omega_s-\omega)a\theta,
\label{e_cap}
\end{equation}
and
\begin{equation}
\Delta E_{free}=2\times\int_{0}^{S}ds \left[\frac{\kappa}{2} \dot{\psi}^2+\Sigma \left( 1-\cos\psi \right)+\omega_s \right],
\label{e_free} 
\end{equation}
where $\dot{\psi}=\frac{d \psi}{d s}$, $\psi (s)$ is the angle between the tangent vector of the membrane $\hat{t}(s)$ and the horizontal axis and $S$ represent the total contour length of the membrane in the membrane segment. 

For the rest of calculations, it will be more convenient to rewrite all expressions in dimensionless units using the bending rigidity $\kappa$ and the cylinder's radius $a$. By defining the following dimensionless parameters
\begin{equation}
\bar{\omega}\equiv\frac{2\omega a^2}{\kappa}; \hspace{5mm} \bar{\omega}_s\equiv\frac{2\omega_s a^2}{\kappa}; \hspace{5mm} \bar{\Sigma}\equiv\frac{2\Sigma a^2}{\kappa},
\label{dimensionless}
\end{equation}
the total energy that is the summation of Eqs. (\ref{e_cap}) and (\ref{e_free}) is written as
\begin{equation}
\bar{E}\equiv \frac{\Delta Ea}{\kappa}=(\bar{\omega}_s-\bar{\omega})\theta+\theta+ \bar{\Sigma} \theta a \left( 1- \frac{\sin \theta}{\theta} \right)+\Delta \bar{E}_{free}.
\label{bar_delta_e}
\end{equation} 
For small angles $\theta \ll 1$, we can write $\dot{\psi}(s)$ as a linear function of $s$ as
\begin{equation}
\dot{\psi}(s)=A+Bs,
\label{psi_dot}
\end{equation}
where $A$ and $B$ are constants. The value of the membrane curvature $\dot{\psi}$ at the membrane-target and membrane-cytoskeleton detachment points ($m$ and $e$, respectively - see Fig. \ref{fig1})  are given by local torque balance equations\cite{Landau_Elasticity} as
\begin{equation}
\dot{\psi}_m=\dot{\psi}(0)=\frac{1}{a}\left(1-\sqrt{\bar{\omega}}\right)\ ,\ \ \dot{\psi}_e=\dot{\psi}(S)=\frac{1}{a}\sqrt{\bar{\omega}_s}\ . 
\label{psi_dot0}
\end{equation}
By integrating Eq. (\ref{psi_dot}), we can calculate the $\psi(s)$ as 
\begin{equation}
\psi(s)=As+\frac{1}{2}Bs^2+C.
\label{psi}
\end{equation}
Combining Eqs. (\ref{psi_dot0})--(\ref{psi}) with the known values of $\psi$ at the two boundaries:
$\psi(0)= \theta$ and $\psi(S)=0$, we obtain:
\begin{equation}
A=\frac{1}{a}(1-\sqrt{\bar{\omega}}),\ BS=\frac{1}{a}(\sqrt{\bar{\omega}_s}-1+\sqrt{\bar{\omega}}),\ C=\theta, 
\end{equation}
and
\begin{equation}
S=\frac{2a\theta}{\sqrt{\bar{\omega}}-1-\sqrt{\bar{\omega}_s}}
\end{equation}

In the small angle limit: $\theta\ll 1$, the energy of the system can be expanded to third order:
\begin{equation}
\bar{E}=A_1(\bar{\omega},\bar{\omega}_s)\theta + A_2(\bar{\omega},\bar{\omega_s},\bar{\Sigma})\theta^{3}.
\label{e_1}
\end{equation}
where $A_1(\bar{\omega}, \bar{\omega}_s )$ and $A_2(\bar{\omega}, \bar{\omega}_s , \bar{\Sigma})$ are 
\begin{equation}
A_1(\bar{\omega}, \bar{\omega}_s )=3+\bar{\omega}_s-\bar{\omega}-\frac{8\sqrt{\bar{\omega}_s}}{3}-2\sqrt{\bar{\omega}}+\frac{8\left( \sqrt{\bar{\omega}}-1 \right)^2}{3\left( \sqrt{\bar{\omega}}-1-\sqrt{\bar{\omega}_s} \right)}
\end{equation}
\begin{equation}
A_2(\bar{\omega},\bar{\omega_s},\bar{\Sigma})=\frac{\bar{\omega}_s+\bar{\Sigma}}{6}+\frac{\bar{\Sigma}}{5\left( \sqrt{\bar{\omega}}-1-\sqrt{\bar{\omega}_s} \right)}\times B(\bar{\omega},\bar{\omega_s})
\end{equation}
and
\begin{equation}
B(\bar{\omega},\bar{\omega_s})=1+\bar{\omega}-2\sqrt{\bar{\omega}}+\frac{8}{3}\bar{\omega}_s+3\sqrt{\bar{\omega}_s}-3\sqrt{\bar{\omega}\bar{\omega}_s}.
\label{e_4}
\end{equation}

By minimizing Eq. (\ref{e_1}) with respect to $\theta$, the wrapping angle is found as
\begin{equation}
\frac{\partial\bar{E}}{\partial \theta}=0 \Rightarrow \theta=\sqrt{\frac{-A_1}{3A_2}}.
\end{equation} 

When $A_1(\bar{\omega}_u,\bar{\omega}_s)$ becomes zero, the transition from the unwrapped to the partial wrapped regimes happens, which implies
\begin{equation}
\bar{\omega}_u=\left( 1+\sqrt{\bar{\omega}_s} \right)^2+\frac{4}{9}\left[ \left( 1+3\sqrt{\bar{\omega}_s} \right)^{3/2}-1  \right]. 
\label{omega-u} 
\end{equation}  
This transition is indicated by a dotted line that separates the unwrapping region from the partial wrapping region in Fig. \ref{fig3} (b).

\subsection{(b) Partial-full wrapped transition}

We first calculate the force (per unit length) acting on the membrane calculating energy changes associated to infinitesimal membrane displacements. With 
$\bm{r}(s)$ the position of the membrane at point $s$, the tangent vector at this point can be defined as $\hat{t}(s)\equiv \partial_s \bm{r}(s)$. In this notation the local curvature of the membrane can be written as
\begin{equation}
\bm{C} \equiv - C(s) \hat{n}(s) = \partial_s \hat{t}(s),
\end{equation} 
where $\hat{n}(s)$ is the normal vector of the membrane at $s$. As a result of the membrane's  deformation amount $\delta \bm{r}(s)$, the tangent vector and the local curvature are changed to $\hat{t}+\delta \hat{t}$ and $\bm{C}+\delta \bm{C}$. After some calculation, we can write
\begin{equation}
\delta \hat{t} = \left[ \partial_s(\delta \bm{r}) \cdot \hat{n} \right] \hat{n}, 
\end{equation}
and
\begin{equation}
\delta \bm{C} =  2 C  \left[ \partial_s(\delta \bm{r}) \cdot \hat{t} \right] \hat{n} + C \left[ \partial_s(\delta \bm{r}) \cdot \hat{n} \right] \hat{t} +\left[\partial_{s,s}(\delta \bm{r}) \cdot \hat{n} \right] \hat{n},
\end{equation}
where $\partial_{s,s} \equiv \frac{\partial^2}{\partial s^2}$. 

The energy of the free part of the membrane considering the contribution of the membrane-cytoskeleton adhesion, can be written as
\begin{equation}
E=\int ds \left[\frac{\kappa}{2} C^2+\Sigma+\omega_s \right]
\end{equation}
and
\begin{equation}
E + \delta E = \int ds^{\prime} \left[ \frac{1}{2} \kappa (\bm{C} + \delta \bm{C} ) ^2 + \Sigma+\omega_s \right],
\end{equation}
where $ds^{\prime}$ is given by $ds^{\prime} = ds \left[ 1 +\hat{t} \cdot \partial_s (\delta \bm{r}) \right]$. After some mathematical steps, one can write
\begin{equation}
\delta E =- \int ds \vec{\mathscr{F}} \cdot \partial _s (\delta \bm{r}),
\label{force1}
\end{equation}
and
\begin{equation}
\vec{\mathscr{F}}(s) = \left[ \frac{1}{2} \kappa C^2(s) - (\Sigma+\omega_s) \right] \hat{t}(s)-\kappa \left[\partial_s C(s) \right] \hat{n}(s),
\label{force2}
\end{equation}
where $\vec{\mathscr{F}}$ is the generalized force per unit length of the cylinder acting on the membrane, as a result of the membrane's deformation by $\delta \bm{r}(s)$. The above equation can be written in terms of $\psi(s)$ as
\begin{equation}
\vec{\mathscr{F}} (s) = \left[ \frac{1}{2} \kappa \dot{\psi}^2 - (\Sigma+\omega_s) \right] \hat{t}(s)+ \kappa \ddot{\psi} \, \hat{n}(s).
\end{equation}

At the equilibrium, the total force acting on each segment of the membrane should be zero, which means that $\vec{\mathscr{F}} (s)$ does not depend on $s$ and is constant. By decomposing of $\vec{\mathscr{F}}$ in $x$ and $y$ directions, we can write
\begin{equation}
\frac{\kappa }{2} \dot{\psi}^2 - (\Sigma+\omega_s)= \mathscr{F}_x \cos \psi + \mathscr{F}_y \sin \psi. \label{force-relation1}
\end{equation}
Using Eq. (\ref{force-relation1}) and the boundary conditions at point $e$, $\psi_e=0$ and $\dot{\psi}_e=\sqrt{\bar{\omega}_s}/a$, the horizontal component of the force can be determined as 
\begin{equation}
\mathscr{F}_x=-\Sigma=-\mathscr{F}_0 \bar{\Sigma} ,
\label{f_x}
\end{equation}
where $\mathscr{F}_0 \equiv \kappa / \left( 2 a^2\right)$. The vertical component of the force can also be calculated using the boundary condition at point $m$, $\psi_m=\theta$ and $\dot{\psi}_m=\frac{1}{a}\left( {1-\sqrt{\bar{\omega}}}\right)$, and Eq. (\ref{force-relation1}) as
\begin{equation}
\frac{\mathscr{F}_y}{\mathscr{F}_0} = \frac{1}{\sin \theta} \left[ \left( 1 - \sqrt{\bar{w}} \right)^2 - \bar{\Sigma} ( 1 - \cos \theta) - \bar{w}_s \right].
\label{f_y}
\end{equation} 
where we have replaced $\theta$ instead of $\alpha_L$ or $\alpha_R$.

\begin{figure}
\centering
\begin{tabular}{cc}
\includegraphics[width=0.5\columnwidth]{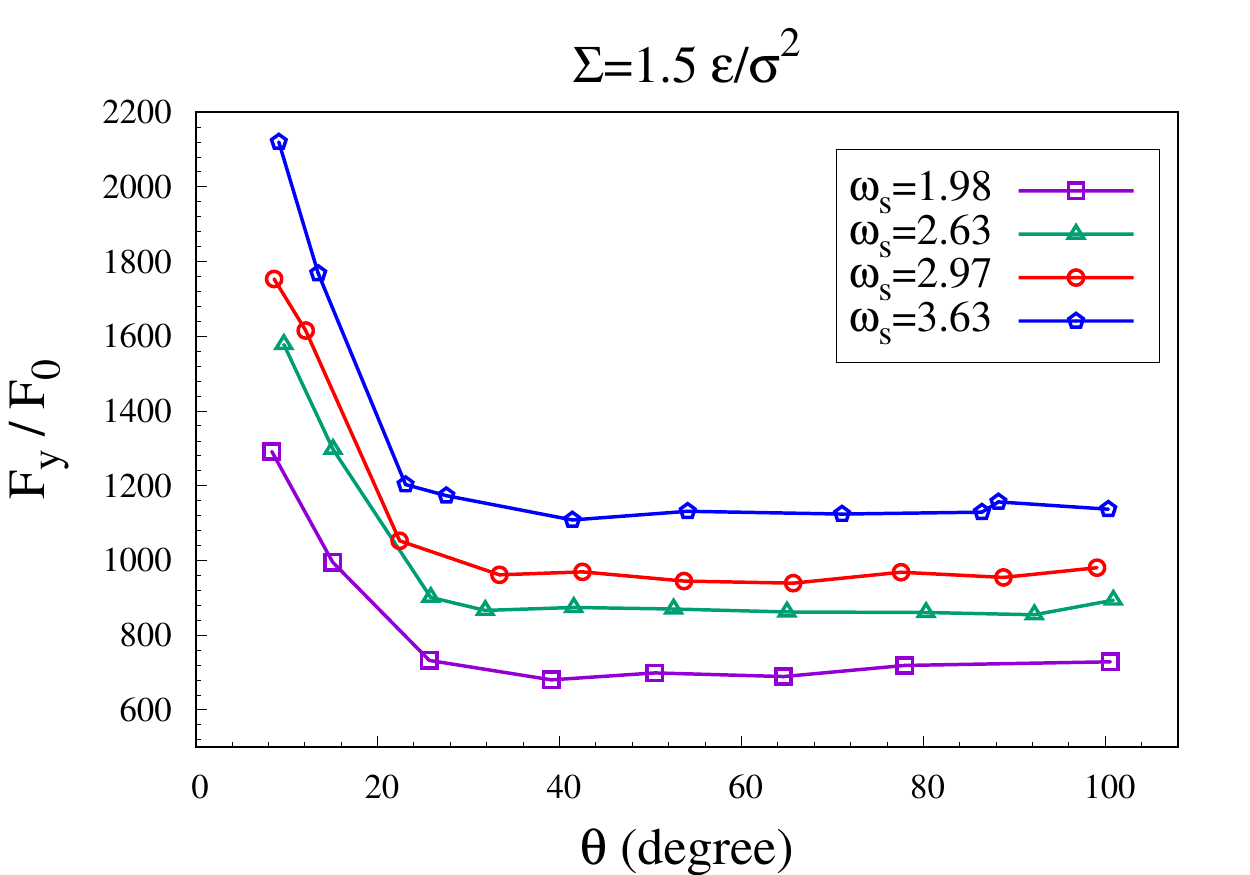}&
\includegraphics[width=0.5\columnwidth]{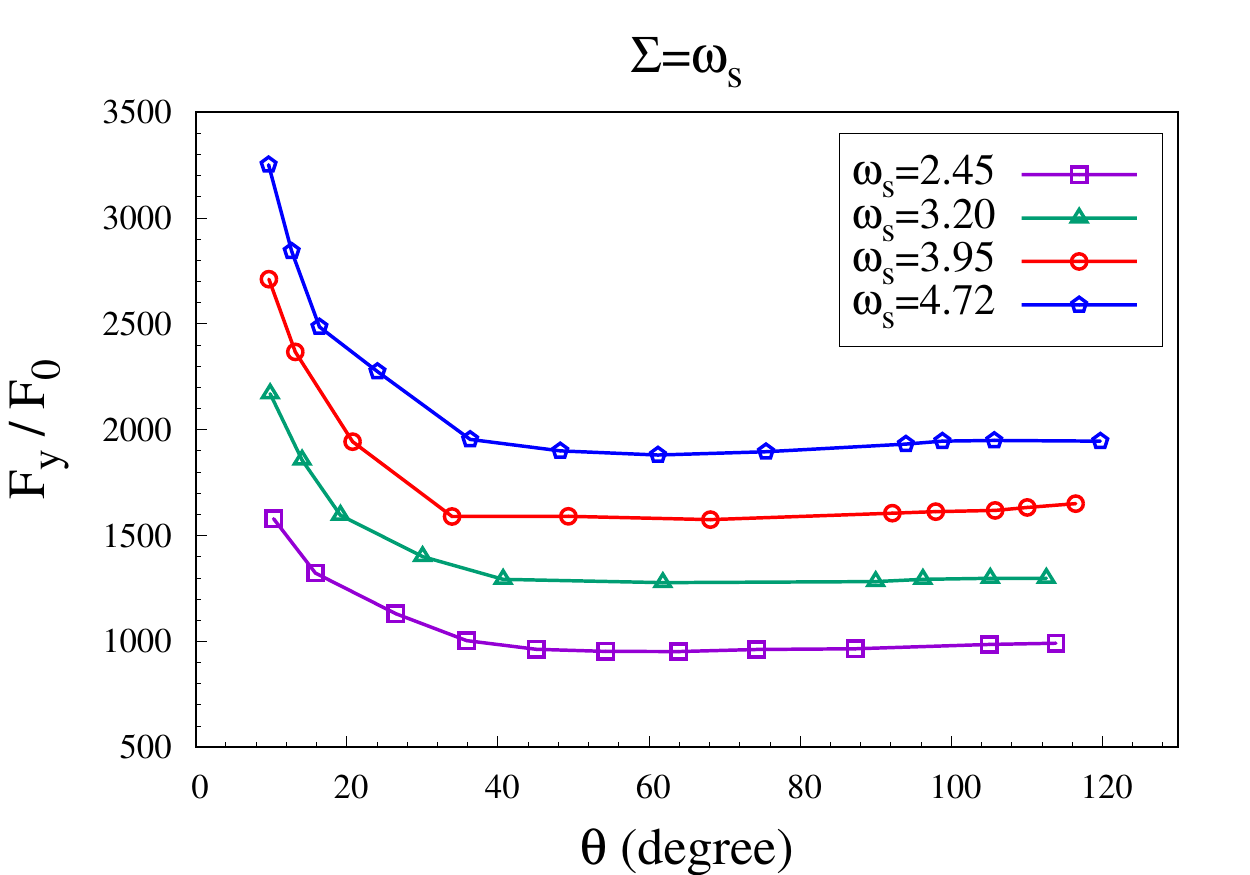}
\end{tabular}
\caption{ The vertical acting force on the membrane, $\mathscr{F}_y$ as a function of the left or right wrapping angle $\theta$, corresponding to the different values of $\omega$ in the simulation.} 
\label{fig-force}
\end{figure}

Fig. \ref{fig3} (a) represents the equilibrium value of the wrapping angle ($\alpha=2\theta$ for symmetric wrapping) as a function of the target binding energy $\omega$. The wrapping angle diverges beyond a value of $\omega$ that depends upon $\omega_s$, which thus correspond to $\frac{\partial \omega}{\partial \theta}=0$. Fig. \ref{fig-force} represents the behavior of $\mathscr{F}_y$ with respect to $\theta$, corresponding to the different values of $\omega$ in the simulation. This force is constant for large enough values of $\theta$ and we can write $\frac{\partial \mathscr{F}_y}{\partial \theta}=0$. Taking the derivation of Eq. (\ref{f_y}) with respect to $\theta$ for given values of $\bar{\omega}_s$ and $\bar{\Sigma}$ and using two above criterion we find
\begin{equation}
\frac{{\mathscr{F}_y}^*}{\mathscr{F}_0}=-\bar{\Sigma} \tan \theta
\label{f_star}
\end{equation}
where ${\mathscr{F}_y}^*$ denotes constant value of the vertical force at the transition point.
Inspection of the membrane shape Fig. \ref{fig2} (see also Fig. \ref{asymmetric}) shows that when the wrapping angle is large, there is a region of the free membrane segment where the angle $\psi$ is constant, named $\psi_0$, which implies $\dot{\psi}|_{\psi_0} = 0$ and $\ddot{\psi}|_{\psi_0} = 0$. As the force $\mathscr{F}$ must be constant through the contour length at the equilibrium, its vertical and horizontal components can be written in terms of $\psi_0$ as
\begin{eqnarray}
\mathscr{F}_y&=&-(\Sigma+\omega_s)\sin \psi_0, \nonumber \\
\mathscr{F}_x&=&-(\Sigma+\omega_s)\cos \psi_0.
\end{eqnarray}
Therefore, we have ${\mathscr{F}_y}^2= \left( \Sigma+\omega_s \right)^2 - {\mathscr{F}_x}^2$ and by substituting $\mathscr{F}_x$ from Eq. (\ref{f_x}) in this equation, we find
\begin{equation}
\frac{\mathscr{F}_y}{\mathscr{F}_0}=\sqrt{\left( \bar{\Sigma}+\bar{\omega}_s \right)^2 - {\bar{\Sigma}}^2}
\label{f_psi_0}
\end{equation}
After substituting Eq. (\ref{f_psi_0}) into Eq. (\ref{f_star}), we can determine the angle $\theta_f$ as
\begin{equation}
\tan \theta_f = - \frac{\sqrt{{\bar{\omega}_s}^2+2\bar{\Sigma}\bar{\omega}_s}}{\bar{\Sigma}},
\label{teta_f}
\end{equation}
where $\theta_f$ denotes the left (right) angle beyond which full wrapping occurs. In the case of $\Sigma=\omega_s$, Eq. (\ref{teta_f}) is converted to $\tan \theta_f = - \sqrt{3}$ and $\theta_f=\frac{2 \pi}{3}$. The wrapping angle $\alpha=2\theta_f$ represented in Fig. \ref{fig3} (a) with the dashed line.
Using Eqs. (\ref{f_y}), (\ref{f_star}), and (\ref{teta_f}), we can find the transition membrane-cylinder binding energy as
\begin{equation}
\bar{\omega}_f=\left( 1+\sqrt{2\left( \bar{\Sigma}+\bar{\omega}_s \right) }\right)^2
\label{omega-f}
\end{equation}
This transition is shown as a dashed line in Fig. \ref{fig3} (b).

\clearpage
\bibliography{Amir_Khosravanizadeh_Endocytosis_Acs_Nano}

\end{document}